\documentclass[a4paper,11pt]{article}

\usepackage[osf]{libertine} 
\usepackage[OT1]{fontenc}
\usepackage{textcomp}
\usepackage[varqu,varl]{inconsolata} 
\usepackage{amsmath,amsthm}
 \usepackage[bottom]{footmisc} 
\usepackage[cal=boondoxo]{mathalfa} 
\usepackage[supstfm=libertinesups,%
  supscaled=1.2,%
  raised=-.13em]{superiors}

\usepackage{amssymb,latexsym,amsmath,amsthm}
\usepackage[hmargin=2.5cm, vmargin=2.5cm]{geometry}
\usepackage{graphicx}
\usepackage[ruled,linesnumbered,vlined]{algorithm2e}
\usepackage{diagbox}
\usepackage{mathptmx} 
\usepackage{verbatim} 
\usepackage{color} 
\usepackage{url} 
\usepackage{graphicx} 
\usepackage{array}
\usepackage{float}
\usepackage{booktabs}
\usepackage{amsmath}
\usepackage{amssymb} 
\usepackage{multirow} 
\usepackage{tabularx} 
\usepackage{polynom} 
\usepackage{threeparttable} 
\usepackage{stackengine}
\usepackage[table,xcdraw]{xcolor}
\newlength\mylen

\usepackage{caption}
\usepackage{subcaption}
\usepackage[numbers]{natbib}
\usepackage{csquotes}
\usepackage{graphicx,epstopdf,array,amsfonts,amssymb,amsmath,amsthm}
\usepackage{pdfpages}
\usepackage{lineno}
\modulolinenumbers[2]
\usepackage[colorlinks=true,linkcolor=blue,breaklinks]{hyperref}%

\usepackage[nameinlink,noabbrev]{cleveref}
\crefname{appendixfigure}{supplementary Figure}{supplementary Figure}
\crefname{appendixtable}{supplementary Table}{supplementary Table}

\allowdisplaybreaks

\graphicspath{{Figure/}}

\usepackage{fancyhdr}
\usepackage[nameinlink,noabbrev]{cleveref}
\crefname{appendixfigure}{supplementary Figure}{supplementary Figure}
\crefname{appendixtable}{supplementary Table}{supplementary Table}

\providecommand{\keywords}[1]{\textbf{\textit{Keywords---}} #1}

\newcommand{\ie}{\textit{i.e.}}

\newcommand{\eqnref}[1]{\eqref{eqn:#1}}

\DeclareMathOperator*{\argmin}{arg\,min}

\newcommand\R{\mathbb{R}}

\newcommand\SO[0]{\mathrm{\mathbf{SO}}}

\newcommand\bth[0]{{\boldsymbol{\theta}}}

\usepackage{tikz,pgfplots,pgfplotstable}
\usepgfplotslibrary{groupplots,fillbetween,colorbrewer,statistics}
\usetikzlibrary{intersections,calc,arrows,matrix,spy,pgfplots.statistics, pgfplots.colorbrewer}
\usepackage{color}
\definecolor{darkred}{rgb}{0.7,0,0}
\definecolor{darkgreen}{rgb}{0,0.5,0}
\definecolor{darkblue}{rgb}{0,0,0.7}
\definecolor{SkyBlue}{rgb}{0.53, 0.81, 0.92}
\pgfplotsset{compat=1.5.1, cycle list/Set1-3}
\usepackage{tikz-3dplot}

\begin{document}
\thispagestyle{plain}
\title{\LARGE \bf
Cryo-forum: A framework for orientation recovery with uncertainty measure with the application in cryo-EM image analysis
}

%
%
%
%
%

\author{
\\
Szu-Chi Chung$^{a,*}$
\\
\\
\\
\\
\\
{$^a$Department of Applied Mathematics, National Sun Yat-sen University, Taiwan}\\
\\
\\
\\
\\
\\
{*Correspondence: \href{mailto: phonchi@math.nsysu.edu.tw }{phonchi@math.nsysu.edu.tw } (S.Z.C.)}
}


\markboth{Target Journal}
{Target Journal}
\maketitle
\newpage
\thispagestyle{plain}

\begin{center}
    \Large
    \textbf{Abstract}
\end{center}
In single-particle cryo-electron microscopy (cryo-EM), the efficient determination of orientation parameters for 2D projection images poses a significant challenge yet is crucial for reconstructing 3D structures. This task is complicated by the high noise levels present in the cryo-EM datasets, which often include outliers, necessitating several time-consuming 2D clean-up processes. Recently, solutions based on deep learning have emerged, offering a more streamlined approach to the traditionally laborious task of orientation estimation. These solutions often employ amortized inference, eliminating the need to estimate parameters individually for each image. However, these methods frequently overlook the presence of outliers and may not adequately concentrate on the components used within the network. This paper introduces a novel approach that uses a 10-dimensional feature vector to represent the orientation and applies a Quadratically-Constrained Quadratic Program to derive the predicted orientation as a unit quaternion, supplemented by an uncertainty metric. Furthermore, we propose a unique loss function that considers the pairwise distances between orientations, thereby enhancing the accuracy of our method. Finally, we also comprehensively evaluate the design choices involved in constructing the encoder network, a topic that has not received sufficient attention in the literature. Our numerical analysis demonstrates that our methodology effectively recovers orientations from 2D cryo-EM images in an end-to-end manner. Importantly, the inclusion of uncertainty quantification allows for direct clean-up of the dataset at the 3D level. Lastly, we package our proposed methods into a user-friendly software suite named {\it cryo-forum}, designed for easy accessibility by the developers.

\vspace{10pt}
\newpage

%
\section{Introduction} \label{introduction}

Proteins, which are large and intricate molecules, are fundamental to all forms of life and perform a plethora of functions within organisms. Historically, scientists have relied on experimental methods such as nuclear magnetic resonance (NMR) and X-ray crystallography to ascertain the structures of proteins. However, these techniques are notoriously labor-intensive, requiring extensive setups and significant trial and error. Cryo-electron microscopy (cryo-EM) has emerged as a promising alternative, becoming the preferred technique for determining 3D protein structures with atomic-level resolution \cite{nakane2020single}. A key advantage of cryo-EM is its ability to analyze conformational mixtures as the molecules are imaged in their near-native states. This capability has proven invaluable since the outbreak of COVID-19 in January 2020, enabling the creation of the first dynamic visualization of the 2019-nCoV Spike trimer structure \cite{wrapp2020cryo} and facilitating the first molecular-level structural analysis of the Omicron variant's spike protein. This analysis has provided crucial insights into how the heavily mutated Omicron variant binds to and infects human cells \cite{mannar2022sars}. Nevertheless, data obtained through cryo-EM are characterized by significant noise, vast dimensions, a large volume of unlabeled data, and high heterogeneity coupled with unknown orientations. These factors complicate the achievement of reliable computational conclusions \cite{singer2020computational}.

It is noted that the process of classifying and refining data collected in a single day can take weeks to process due to the complexity involved in estimating 3D orientations and the necessity for human intervention in the clean-up process. In this study, we build on the concepts presented in \cite{banjac2021learning}, which employed contrastive learning and neural networks to estimate the distances of orientations between projections. We believe incorporating pairwise relationships can add regularization to orientation learning, thus enabling us to alleviate the issues posed by high noise levels. Furthermore, utilizing neural networks that carry out amortized inference on orientation can substantially reduce the processing time. Lastly, we suggest a novel uncertainty measure and reconstruction pipeline to assess the reliability of our estimates and speed up the entire image analysis process. The contributions of this paper can be summarized as follows:

\begin{enumerate}
\item \textbf{Uncertainty estimation for orientation estimation}: Evaluating the reliability of the network's predictions is critical, particularly given the substantial presence of outliers and contaminants in the cryo-EM dataset. Uncertainty estimation is vital in quantifying the dependability of the network's predictions and can aid us in filtering particles during data cleaning. In this study, we introduce uncertainty measures that can serve as proxies for estimating testing errors. Additionally, we propose a strategy, based on the uncertainty measure, to clean up the dataset directly at the 3D level, which, as we demonstrate, can lead to a more accurate 3D reconstruction. This advancement could potentially decrease the time spent on dataset clean-up using traditional 2D classification methods.

\item \textbf{Model design and generalization capability}: Another essential concern is the model's ability to generalize effectively. In this study, we systematically assess the network's generalization capabilities. Specifically, we suggest using distance learning as an auxiliary loss to regularize the learning process and explore the potential of different components in the neural network. This comprehensive study can guide us in understanding the design choices made when leveraging a neural network for amortized inference. Furthermore, it addresses the current gap in the design of the encoder network in the generative model, an emerging framework in 3D reconstruction \cite{levy2022amortized}. Ultimately, this analysis may pave the way to obtain a pre-trained model, thereby further accelerating the process of orientation estimation.
\end{enumerate}

By addressing these critical issues, this paper aims to make significant contributions to the ongoing efforts to leverage neural networks for orientation estimation in cryo-EM \cite{banjac2021learning,lian2022end} and to improve the design of encoders in current generative models \cite{levy2022amortized,nashed2021cryoposenet,levy2022cryoai}. The rest of this paper is organized as follows. In \Cref{related_works}, we first review the related work on orientation estimation, followed by an examination of the related work on generative models that use 3D orientation in their autoencoder frameworks. In \Cref{methods}, we present the design of the proposed framework and provide insights into each component. In \Cref{experiments}, we offer numerical results to compare different design choices in the methods and demonstrate the superior performance of the proposed framework. Finally, in \Cref{discussion}, we discuss the potential of the methodology and draw conclusions.

\section{Related works} \label{related_works}
Following the review by \cite{levy2022cryoai}, we classify the previous work on orientation regression for 2D projections from 3D volume $V$ into two main inference categories: non-amortized and amortized. Non-amortized inference comprises methods in which the posterior distribution of the orientations $\theta$, denoted as $p(\theta_i|I_i, V)$, is calculated independently for each individual image $I_i$. Notable strategies within this category include projection-matching methods \cite{tang2007eman2,grant2018cistem} and Bayesian formulations \cite{scheres2012relion}. For instance, in the RELION framework \cite{scheres2012relion}, the posterior distributions over the orientations are computed for each image during the expectation step, and all the frequency components of the volume are updated in the subsequent maximization step. However, this process is computationally demanding due to the iterative nature of the expectation-maximization algorithm. Alternatively, cryoSPARC \cite{punjani2017cryosparc} proposes collectively performing Maximum A Posteriori (MAP) optimization using Stochastic Gradient Descent (SGD) to optimize the volume $V$ alongside branch-and-bound algorithms for estimating the orientations $\theta_i$. While cryoSPARC's gradient-based optimization scheme for $V$ mitigates the computational burden associated with RELION's updates in the maximization step, it still necessitates estimating the orientation for each image by aligning every 2D projection $I_i$ with the estimated 3D volume $V$, which can also be computationally intensive.

Amortized inference techniques, on the other hand, learn a parameterized function, $q(I_i)$, that approximates the posterior distribution of the orientations, $p_{\xi}(\theta_i|I_i, V)$. These methods simplify the process by eliminating the orientation matching step, which is the main computational bottleneck in non-amortized methods, albeit at the expense of optimizing the parameter $\xi$. Lian et al. \cite{lian2022end} investigated the feasibility of using a convolutional neural network to approximate the relationship between cryo-EM images and orientations. However, their approach's effectiveness depends on the availability of an accurate 3D initial model and the selection of hyperparameters to balance the re-projection and orientation estimation loss. Banjac \cite{banjac2021learning} noted that directly regressing on orientation parameters can be unstable and introduced a two-phase strategy. The first phase uses contrastive learning to estimate pairwise distances between orientations of particle images, while the second phase recovers the estimated orientations from these distances. Despite this, the method does not yield satisfactory results, and the two-step process extends the computational time due to the orientation recovery procedure. CryoPoseNet \cite{nashed2021cryoposenet} and CryoAI \cite{levy2022cryoai} introduced an autoencoder framework with a specially designed decoder capable of carrying out homogeneous reconstruction with the volume randomly initialized, thus eliminating the need for the 3D initial model. However, as these methods estimate both the volume and orientation simultaneously, the loss function requires careful design to avoid local minima. Lastly, all the aforementioned approaches do not account for the network's uncertainty and overlook the discussion on the generalization problem.


\section{Methods and framework} \label{methods}
\begin{figure}[!b]
\centering
\includegraphics[width=1\textwidth]{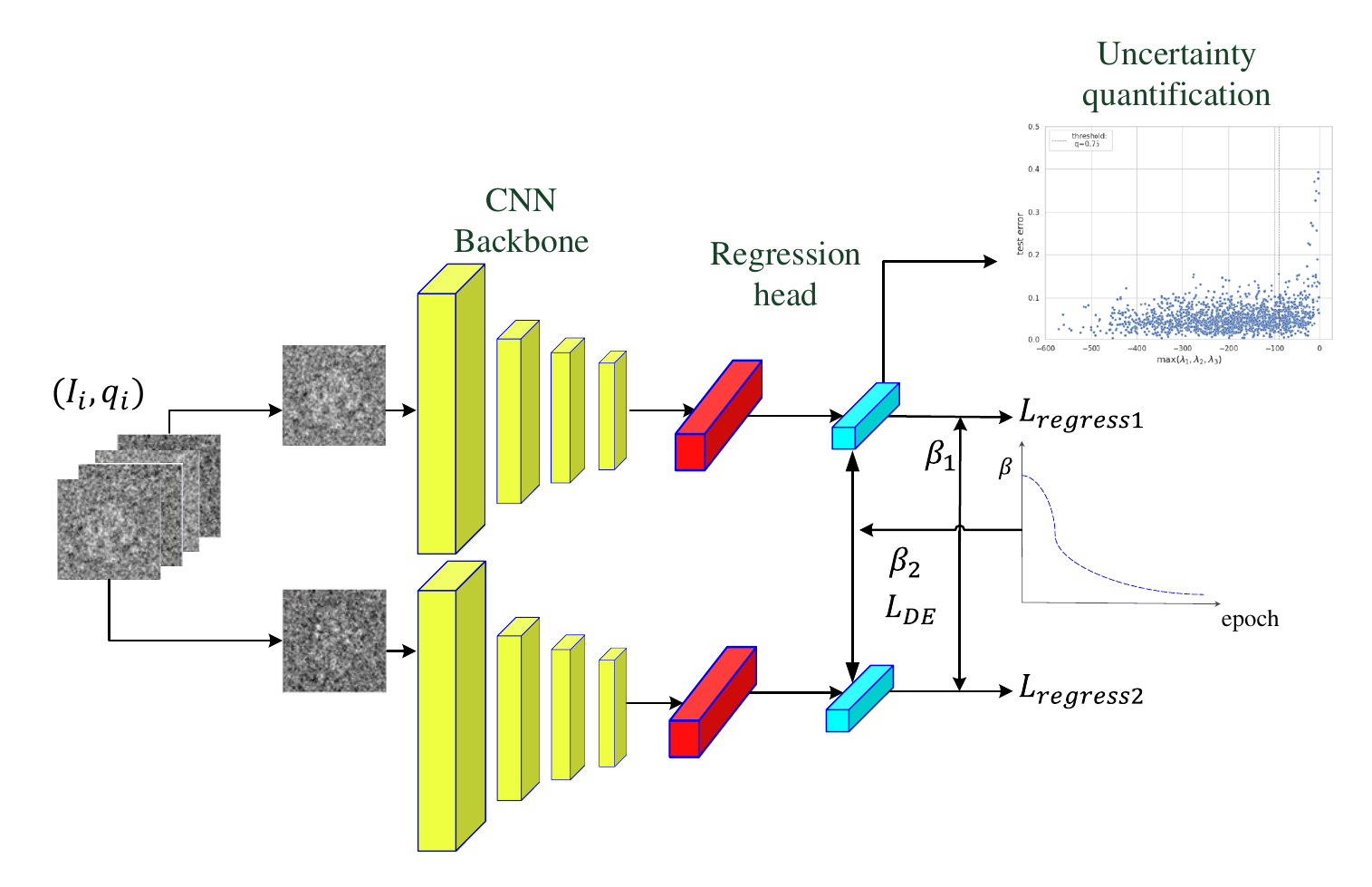} 
\caption{ {\it Cryo-forum} framework: The network is composed of two Convolutional Neural Network (CNN) backbones, each coupled with a regression head. Importantly, these two networks share the same weights. The backbones are used to encode the image into an embedding space (illustrated in red), after which a fully connected layer is employed to translate the representation into orientation and to compute the regression loss. Additionally, a distance loss is calculated, which measures the discrepancy between the distance in the final representation space (illustrated in blue) and the distance corresponding to the ground truth orientation. A curriculum is devised to manage the weighting of these two losses during the network training process. Ultimately, uncertainty is extracted from the final representation space to gauge the confidence level of the estimation and to filter out the contaminates.}
\label{fig:framework}
\end{figure}

In this section, we delve into the key components of our framework called {\it cryo-forum}. The graphical illustration of our framework is presented in \Cref{fig:framework}.

\subsection{Representation of orientations}

The 3D orientation of a particle to the plane of the microscope's detector is essentially a rotation compared to a reference orientation. The set of all 3D rotations corresponds to $\SO(3)$, representing the group of $3 \times 3$ orthogonal matrices with a determinant of 1 under the matrix multiplication operation. A 3D rotation matrix, denoted as $\mathbf{R}_{\bth} \in \SO(3)$, can be decomposed into the product of three independent rotations. For example, $\mathbf{R}_{\bth} = \mathbf{R}_{\psi} \mathbf{R}_{\theta} \mathbf{R}_{\phi}$, where $\bth = (\psi,\theta,\phi) \in [0,2\pi) \times [0,\pi) \times [0,2\pi)$ represent the Euler angles following the $ZYZ$ convention, a commonly used parameterization in cryo-EM \cite{sorzano2014interchanging}. Despite their compact representation of orientation (three values for three degrees of freedom), Euler angles are burdened by a topological constraint, wherein a continuous mapping between the Euler angle and $\SO(3)$ is nonexistent. This attribute makes their optimization through gradient descent problematic \cite{zhou2019continuity}. On the other hand, optimizing 3D rotation matrices, which consist of nine elements, would require imposing computationally burdensome constraints (like orthogonality and a determinant of 1) to limit the degrees of freedom to three. Furthermore, calculating the distance between orientations is not straightforward when using Euler angles and becomes computationally demanding when derived from 3D rotation matrices \cite{huynh2009metrics}. As a solution to these challenges, we advocate for the adoption of unit quaternions, as demonstrated in \cite{banjac2021learning,lian2022end}. 



Quaternions $q \in \mathbb{H}$ are an extension of complex numbers \footnote{The algebra $\mathbb{H}$ is similar to the algebra of complex numbers $\mathbb{C}$, except for multiplication being non-commutative.} of the form $q = a + b\mathbf{i} + c\mathbf{j} + d\mathbf{k}$ where $a,b,c,d \in \R$. In addition, unit quaternions $q \in \mathbb{S}^3$, where $\mathbb{S}^3 = \big\{ q \in \mathbb{H}: \lvert q \rvert = 1 \big\}$ is the 3-sphere 
, concisely represent a rotation of angle $\theta$. Unlike Euler angles, $\mathbb{S}^3$ is isomorphic to the universal cover of $\SO(3)$.
Hence, the distance between two orientations, \ie, the length of the geodesic between them on $\SO(3)$, is well defined and can be efficiently calculate:
\begin{equation}
    \begin{aligned}
        d_q &: \mathbb{S}^3 \times \mathbb{S}^3 \rightarrow [0,\pi], \\
        d_q(q_i, q_j) &= 2 \arccos \left( \left| \langle q_i, q_j \rangle \right| \right),
    \label{eqn:distance:orientations}
    \end{aligned}
\end{equation}
where $\langle \cdot, \cdot \rangle$ is the inner product.
The absolute value $\left| \cdot \right|$ ensures that $d_q(q_i, q_j) = d_q(q_i, -q_j)$ as $q$ and $-q$ represent the same orientation because $\mathbb{S}^3 \rightarrow \SO(3)$ is a two-to-one mapping. However, as noted in \cite{peretroukhin2020smooth},  a standard unit quaternion parameterization does not satisfy an important continuity property that is essential for learning arbitrary rotation targets. To address these issues, authors in \cite{levinson2020analysis} develop alternative rotation representations that satisfy this property and lead to better network performance. All of these representations, however, are point representations and do not quantify network uncertainty. 

In this work, we proposed a representation based on \cite{peretroukhin2020smooth}. Specifically, we represent the rotation as the set of real positive semidefinite $4\times4$ matrices with a non-repeated minimum eigenvalue. The idea is that if we would like to find the spatial rotation that aligns associated two vector measurements, then we need to solve the following Quadratically-Constrained Quadratic Program (QCQP) \cite{yang2019quaternion}:
\begin{equation*}
    \begin{aligned}
        \min_{q \in R^4} &: q^TAq \\
        \text{subject to}   \quad & q^Tq=1,
    \end{aligned}
\end{equation*}
Here, $A$ is constructed from the two associated vectors we aim to align. In the context of cryo-EM, considering that all particles are projections of the same 3D structure in a common coordinate system (assuming homogeneous reconstruction), we can perceive one of the vectors as a fixed reference and the 3D orientation of the input 2D projection is exactly the relative orientation to the reference vector. Specifically, we address the following problem, as described in \cite{peretroukhin2020smooth}:

\begin{equation}
    \begin{aligned}
        \min_{q \in R^4} &: q^TA(\theta)q \\
        \text{subject to}   \quad & q^Tq=1,
    \label{eqn:param_qcqp}
    \end{aligned}
\end{equation}
In this context, we impose the additional constraint that $A$ should be a positive semidefinite (PSD) matrix so that it can be related to the Bingham Distribution later. It is sufficient to parameterize the triangular part of a symmetric matrix, and so we encode $A$ as:

\begin{equation}
\begin{aligned}
A &= LL^T \\
L &=\begin{bmatrix}
\theta_1 & \theta_2 & \theta_3 & \theta_4\\
0 & \theta_6 & \theta_7 & \theta_8 \\
0 & 0 & \theta_{10} & \theta_9\\
0 & 0 & 0 & \theta_5\\
\end{bmatrix}
 \end{aligned}
\end{equation}

In practice, we employ amortized inference to learn a parameterized function that maps each particle image to the embedding space formed by $A$, thereby avoiding the need to calculate the matrix for each image explicitly. A neural network is used to encode a particle image into a ten-dimensional space. Then, the smallest eigenvalue of $q^TA(\theta)q$ serves as the solution to the above problem. The representation can be proven to be smooth using the same argument as in \cite{peretroukhin2020smooth}, and the output is simply a four-dimensional unit quaternion. Furthermore, both the mapping and the eigenvalue decomposition are differentiable, and the gradient in backpropagation can be calculated using an auto-differentiation engine like TensorFlow \cite{abadi2016tensorflow}. Finally, it should be noted that the unit quaternion representation in the output is useful in our study, as several metrics, such as the distance between two quaternions, are well-defined \cite{hu2020unit}.

\subsection{Distance learning}

One approach for efficient orientation estimation is to employ a regressor, $f_w$, to directly establish a mapping from projections $I_i$ to unit quaternions, denoted as $\widehat{q_i} = f_w(I_i)$. However, the estimation may be unstable, as noted by \cite{banjac2021learning}. In previous works, the 3D information is employed to stabilize the learning \cite{lian2022end,nashed2021cryoposenet,levy2022cryoai}, which may require a complicated and expensive setup. Here, we take a different route that incorporates the contrastive learning approach to address this problem. The approach was first introduced in \cite{banjac2021learning}, which employs a Siamese Network (SNN) \cite{chopra2005learning} to bolster the training process. Additionally, we find that incorporating the information about pairwise distances and judiciously scheduling the weights of different components within the loss function can further enhance the training, as elucidated in subsequent sections.

Our learning paradigm is formalized as follows: Given two projections $(I_i, I_j)$ and their orientations $(q_i, q_j)$ represented using unit quaternion, our framework learns the pairwise projection distance $d_I$ as follows:
\begin{equation*}
    \widehat{d_I} = \argmin_{d_I} L_\text{DE},
    \quad \text{where} \quad
    L_\text{DE} = \sum_{i,j} \left| d_I\big(I_i,I_j\big) - d_q\big(q_i,q_j\big) \right|
    \label{eqn:distance-learning}
\end{equation*}
where $d_q$ is defined in~\eqnref{distance:orientations} and $d_I$ is parameterized as the Siamese neural network:
\begin{equation*}
    d_I(I_i, I_j) = d_q(f_w(I_i), f_w(I_j)),
\end{equation*}
where $f_w$ is a convolutional neural network with weights $w$ that is trained to extract the most relevant features from a projection $I_i$. In addition, we also add a regression head to map the feature vectors to the true orientations. The regression loss for the regression head is explicitly added to the loss function. Therefore, our loss function can be written as follows:
\begin{equation*}
    L_\text{regress} = d_q(f_w(I_i), q_i) 
\end{equation*}
\begin{equation}
    L_\text{FDE} =\beta_1 (d_q(f_w(I_i), q_i) + d_q(f_w(I_j), q_j)) + \beta_2 L_\text{DE}
    \label{eqn:full_loss}
\end{equation}

In constructing the architecture for $f_w$, we impose constraints on the function space by incorporating our prior knowledge. Specifically, we employ a convolutional architecture to achieve shift invariance – ensuring that a spatial shift does not alter the estimated distances and orientations. Size invariance, which enables the handling of projections $I$ of different sizes while generating fixed-size representation features, is attained through a final global average pooling layer. The invariance to noise or contrast transfer function (CTF) is achieved by employing data augmentation, which trains the model on perturbed projections.


\subsection{Preprocessing and blurring layer}

\begin{figure}[hbt!]
\centering
    \includegraphics[width=4.2in]{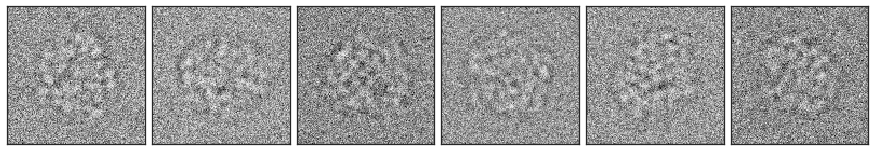}
      \includegraphics[width=4.2in]{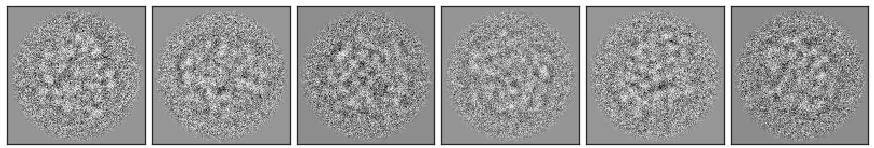}
        \includegraphics[width=4.2in]{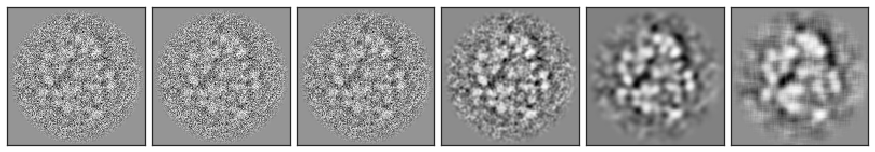}
          \includegraphics[width=4.2in]{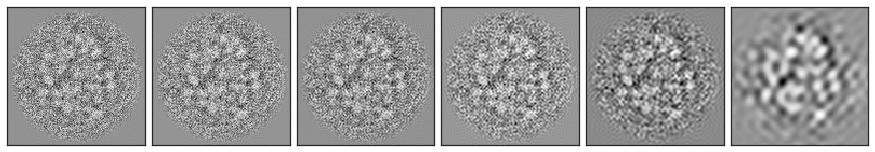}
\caption{An illustration of the preprocessing results. The first row shows the six randomly selected cryo-EM images. The second row shows the images after masking and standardization. The third row shows the first image in the second row after the Gaussian filter with different kernel sizes. The fourth row shows the first image in the second row after the low pass filter with different cutoff frequencies.}
\label{fig:prepro}
\end{figure}

To reduce the computation cost, the particle images are initially resized to a resolution of 128 pixels. A circular mask is then applied to the input images, which helps minimize the interference from background noise. The masked image is standardized and forwarded into the network.  The design of the first layer of the network draws inspiration from \cite{chung2020pre}, which suggests that low-resolution information is critical for image alignment in the early stages. Therefore, we implement a blurring layer that filters each input image using a set of low-pass filters with varying cutoff frequencies. The filtered images are then concatenated with the original image along the depth dimension to construct a multi-frequency representation of the initial image. We hypothesize that the network can first utilize the low-frequency component for early coarse alignment and switch to the high-frequency component for finer alignment in later iterations so that it will contribute to a more reliable orientation estimation. A similar technique utilizing Gaussian filters was recently employed in \cite{levy2022cryoai}. The preprocessing results and the filtering outcomes can be observed in \Cref{fig:prepro}. 

\subsection{Network design}

The backbone of our Siamese neural network is structured as follows: After the blurring later, the filtered image collection is input into two convolution blocks encompassing convolution, activation, and max-pooling layers. The kernel size of these two layers is 7 and 5, and the channel size is 32 and 64, respectively. This is followed by four consecutive DoubleConv layers, producing 64, 128, 256, and 1,024 channels. Each DoubleConv layer comprises two convolutional layers with kernel sizes of 3 and a max-pooling layer that halves the height and width of each image. This architecture, reminiscent of VGG, is renowned for its feature extraction efficacy across various applications. At this stage, assuming an input image size of 128 pixels, the feature vector possesses dimensions of $2^2 \times 1024$. This feature vector is then channeled into a global pooling layer, enabling $f_w$ to adapt to any image size and yield a feature vector with 1,024 dimensions. After experimenting with various global pooling operations, we observed that the generalized mean pooling \cite{radenovic2018fine}, defined as:

\begin{equation*}
    \textbf{f} = [f_1^{(g)}, \ldots, f_K^{(g)}], \quad f_k^{(g)} = \left(\frac{1}{|X_k|}\sum_{x \in X_k}x^{p}\right)^{1/p}
    \label{eqn:GeM}
\end{equation*}
is the most effective in practice. Here, $\textbf{f}$ represents the final feature vector with $K$ dimensions, and $||$ denotes the cardinality. Each element of the vector is calculated by computing the generalized mean of the elements $x$ within each channel $X_k$ with different $p$ determined by the data. It's noteworthy that as $p$ approaches infinity, this becomes global max pooling, and for $p=1$, it reduces to global average pooling. Finally, a fully connected layer is employed to map the pooled feature vector to a ten-dimensional space.

For the selection of the activation function, we choose the Parametric ReLU (PReLU), which provides more accurate modeling \cite{he2015delving} and is defined as follows
\begin{equation*}
    f(x_i) = 
    \begin{cases}
    x_i, & \text{if } x_i > 0\\
    \alpha_i x_i, & \text{if } x_i \leq 0
    \end{cases}
    \label{eqn:Prelu}
\end{equation*}
where $x_i$ is the input to the activation function on the $i$th channel and $\alpha_i$ is the learned coefficient controlling the slope of the negative part. Notice that when $\alpha_i=0$, it reduces to the standard ReLU function. HE initialization \cite{he2015delving} is used to initialize the weights of all the layers. Since we observe overfitting when training the above neural network, dropout layers and $l_2$ regularization are added to the design. Finally, batch normalization (BN) is included to enhance the generalization ability. The network architecture is listed in \Cref{tbl:arch}.

\subsection{Curriculum learning and sampling scheme }

For the training process, we implement a curriculum that governs the weights of $\beta_1$ and $\beta_2$, which are by default set as $\beta_1=(\frac{i}{l})^{1/2}/2$ and $\beta_2=1-(\frac{i}{l})^{1/2}$ for the $i$-th iteration, where $l$ denotes the total number of training iterations. The proposed learning approach is depicted in \Cref{fig:framework}. The idea is that we encourage the network to learn the geometry of $\SO(3)$ through pairwise distances in the early iterations, while the actual orientation estimation will gradually dominate the training in the later iterations. 

Additionally, the selection of the training pairs in contrastive learning will greatly affect the learning process. Since calculating the sum over $N^2$ pairs is computationally prohibitive for cryo-EM datasets, where $N$ is the number of images and is typically in the order of $10^5$ projections, two schemes are devised as shown in \Cref{fig:histograms}. The first approach is that we randomly sample $1\%$ of the image pairs and sum them in order to minimize the loss \eqnref{full_loss} via SGD applied to small batches of pairs. The second approach implements stratified sampling that equalizes the number of training pairs with different distances. Specifically, we first sample a large number of pair of indices and build the histogram of the distances between training pair. In the second step, we find the minimum number of training pairs along each bin and randomly discard the training pairs in the other bin until the number of training pairs in each bin is equalized. Empirically, we find that both approaches work well in the synthetic dataset, but the latter tends to have slightly better performance in the real dataset, which may be due to the problem of preferred orientation existing in the real dataset.

For the optimizer, the Adam optimizer is employed with a minibatch size set to 256 for optimization purposes. Furthermore, diverging from prior studies that make use of a fixed learning rate coupled with early stopping or performance scheduling based on validation loss, we utilize one-cycle learning in this study. One-cycle learning aids in achieving super-convergence and mitigates the risk of overfitting \cite{smith2019super}. The scheduling can be visualized in  \Cref{fig:1cycle}.

\subsection{The uncertainty measure and the filtering strategy}
As elaborated in \cite{peretroukhin2020smooth}, the ten-dimensional representation $A$ defines the Bingham distribution over unit quaternions. Recall that the probability density function of the Bingham distribution for unit quaternions is given as:

\begin{equation}
    p_{Bingham}(x;D,\Lambda) =\frac{1}{Z}\exp(x^TD\Lambda D^Tx) = \frac{1}{Z}\exp(\sum_{i=1}^3 \lambda_i(d_i^Tx)^2)
    \label{eqn:Bingham}
\end{equation}
Here, $x$ belongs to a unit sphere $\mathbb{S}^3$, $Z$ is the normalization constant, and $D$ is an orthogonal matrix formed by three unit vectors $d_i$ and a fourth mutually orthogonal unit vector $d_4$. $\Lambda$ is a diagonal matrix with diagonal elements $(\lambda_1, \lambda_2, \lambda_3, 0)$ and $\lambda_1 \le \lambda_2 \le \lambda_3 \le 0$. Each $\lambda_i$ governs the dispersion of probability mass along the direction specified by $d_i$ and is termed the dispersion coefficient. In addition, a smaller magnitude of $\lambda$ corresponds to a larger spread. Finally, $d_4$ represents the mode of the distribution, and $\lambda_i$ are the eigenvalues of $D\Lambda D^T$. To establish a relationship with the dispersion coefficients and the matrix $A$, we can compute the non-zero eigenvalues of $-A+\lambda_1^AI$ \footnote{The transformation here ensures that the maximum eigenvalue of $-A+\lambda_1^AI$ equals zero to align with the eigenvalues of $D\Lambda D^T$.}, where $\lambda_i^A$ denotes the eigenvalues of $A$ in ascending order. The matrix $-A+\lambda_1^AI$ defines the density $p_{Bingham}(x;D,\Lambda+\lambda_1^AI)$, which is equivalent to $p_{Bingham}(x;D,\Lambda)$ according to \cite{darling2016uncertainty,glover2013tracking}. Consequently, we have ${\lambda_1, \lambda_2, \lambda_3}={-\lambda_4^A+\lambda_1^A, -\lambda_3^A+\lambda_1^A, -\lambda_2^A+\lambda_1^A}$.

Based on the above interpretation, it is possible to define statistics that quantify the uncertainty of the network. Consequently, we define two statistics:

\begin{equation}
    \Lambda_{max} = \max(\lambda_1, \lambda_2, \lambda_3) = \max(-\lambda_4^A+\lambda_1^A, -\lambda_3^A+\lambda_1^A, -\lambda_2^A+\lambda_1^A)
    \label{eqn:max_lambda}
\end{equation}

\begin{equation}
    tr(\Lambda)=\sum_i^3\lambda_i=3\lambda_1^A-\lambda_2^A-\lambda_3^A-\lambda_4^A
    \label{eqn:trace}
\end{equation}
The first statistic \eqnref{max_lambda} measures the largest dispersion along a given direction, while the second statistic \eqnref{trace} quantifies the overall dispersion for all directions. To exclude outliers or particle images with low confidence, a threshold on the above statistics can be determined based on quantiles over the testing data, retaining samples with higher magnitudes of statistics that represent higher confidence.

\section{Experiment Results} \label{experiments}
To demonstrate the design choice of different components inside our framework, we first apply our methodology to synthetic cryo-EM data, which is prepared as follows. Firstly, we downloaded the RELION \cite{scheres2012relion} benchmark dataset, \href{https://www.ebi.ac.uk/pdbe/emdb/test_data.html}{E. coli 70S ribosome}, which contains 10,000 particle images each of size $130 \times 130$ pixels. The dataset is divided into two parts, with the first 5,000 images representing one structural conformation and the second 5,000 images representing a different conformation. Next, we employed cryoSPARC \cite{punjani2017cryosparc} to generate a 3D density map from the first set of 5,000 images, corresponding to the ribosome bound with an elongation factor (EF-G). This density map was resized to $128 \times 128$ pixels with a pixel size of 2.86 \AA{}. Subsequently, 5,000 distinct 2D images of size $128 \times 128$ pixels were generated by projecting the 3D density map in uniformly spaced orientations according to the HEALPix framework of RELION. The distribution of the orientation sampling is illustrated in \Cref{fig:sampling}. In addition, random shifts along $x$ and $y$ directions are applied to each particle image to model potential residual after particle picking. Following this, each image was multiplied with the CTF in Fourier space. The CTF was randomly sampled from the experimental dataset. Lastly, independent and identically distributed (i.i.d.) Gaussian noise, $N\left(0, \sigma^2I\right)$, with value of $\sigma^2$, was added to the images to create datasets with a signal-to-noise ratio (SNR) of 0.1. This process results in a synthetic dataset that closely resembles real cryo-EM data, allowing us to evaluate the performance of our methodology in a controlled setting.

We split the projections into training, validation, and test subsets. The number of images in each split is 50$\%$, 17 $\%$ and 33$\%$ of the whole dataset, respectively. The error is calculated using 
\begin{equation}
    \begin{aligned}
        d_q(q_i, \widehat{q_i}) &= 2 \arccos \left( \left| \langle q_i, \widehat{q_i} \rangle \right| \right),
    \end{aligned}
\end{equation}
where $q_i$ and $\widehat{q_i}$ are the true and predicted unit quaternion of the $i$-th image, respectively. We train the network for 50 epochs with a learning rate of 0.001 and report the error's median \footnote{median is chosen here instead of mean since there are a portion of outliers exist for the network's prediction.} and variance over five runs on the training, validation and testing dataset. Finally, all the experiments are performed on a single Tesla V100 GPU.

\begin{table}[ht!]
\captionsetup{justification=centering, singlelinecheck=false}
\centering
\caption{Error associated with different strategy. The results show the median testing error and variance calculated on five independent runs.}
\begin{tabular}{c|ccc}
\hline
                                                                               & \multicolumn{3}{c}{Rotational representation}                                                                                                                                                                                      \\ \hline
                                                                               & \multicolumn{1}{c|}{Training Error}                                              & \multicolumn{1}{c|}{Validation Error}                                            & Testing Error                                                \\ \hline
Unit Quaternion                                                                & \multicolumn{1}{c|}{\begin{tabular}[c]{@{}c@{}}0.0217 \\ (0.0006)\end{tabular}} & \multicolumn{1}{c|}{\begin{tabular}[c]{@{}c@{}}0.1071 \\ (0.0021)\end{tabular}} & \begin{tabular}[c]{@{}c@{}}0.01081 \\ (0.0015)\end{tabular} \\ \hline
6D Representation                                                              & \multicolumn{1}{c|}{\begin{tabular}[c]{@{}c@{}}0.0196\\ (0.0007)\end{tabular}}  & \multicolumn{1}{c|}{\begin{tabular}[c]{@{}c@{}}0.0943\\ (0.0022)\end{tabular}}  & \begin{tabular}[c]{@{}c@{}}0.0981 \\ (0.0018)\end{tabular}  \\ \hline
Modified QCQP                                                                  & \multicolumn{1}{c|}{\begin{tabular}[c]{@{}c@{}}0.0162\\ (0.0005)\end{tabular}}  & \multicolumn{1}{c|}{\begin{tabular}[c]{@{}c@{}}0.0696\\ (0.0018)\end{tabular}}  & \begin{tabular}[c]{@{}c@{}}0.0742 \\ (0.0016)\end{tabular}  \\ \hline
                                                                               & \multicolumn{3}{c}{Training strategy}                                                                                                                                                                                              \\ \hline
Single branch                                                                  & \multicolumn{1}{c|}{\begin{tabular}[c]{@{}c@{}}0.0430 \\ (0.0011)\end{tabular}} & \multicolumn{1}{c|}{\begin{tabular}[c]{@{}c@{}}0.1119\\ (0.0024)\end{tabular}}  & \begin{tabular}[c]{@{}c@{}}0.1166\\ (0.0031)\end{tabular}   \\ \hline
Siamese Network                                                                & \multicolumn{1}{c|}{\begin{tabular}[c]{@{}c@{}}0.0131\\ (0.0001)\end{tabular}}  & \multicolumn{1}{c|}{\begin{tabular}[c]{@{}c@{}}0.0750\\ (0.0013)\end{tabular}} & \begin{tabular}[c]{@{}c@{}}0.0787\\ (0.0006)\end{tabular}   \\ \hline
\begin{tabular}[c]{@{}c@{}}Siamese Network \\ with auxiliary loss\end{tabular} & \multicolumn{1}{c|}{\begin{tabular}[c]{@{}c@{}}0.0162\\ (0.0005)\end{tabular}}  & \multicolumn{1}{c|}{\begin{tabular}[c]{@{}c@{}}0.0696\\ (0.0018)\end{tabular}}  & \begin{tabular}[c]{@{}c@{}}0.0742 \\ (0.0016)\end{tabular}  \\ \hline
                                                                                         & \multicolumn{3}{c}{Blurring layer}                                                                                                                                                                                             \\ \hline
                                                                                         & \multicolumn{1}{c|}{Training Error}                                             & \multicolumn{1}{c|}{Validation Error}                                           & Testing Error                                              \\ \hline
None                                                                                     & \multicolumn{1}{c|}{\begin{tabular}[c]{@{}c@{}}0.0166\\ (0.0009)\end{tabular}}  & \multicolumn{1}{c|}{\begin{tabular}[c]{@{}c@{}}0.0744 \\ (0.0057)\end{tabular}} & \begin{tabular}[c]{@{}c@{}}0.0789\\ (0.0048)\end{tabular}  \\ \hline
Gaussian filter                                                                          & \multicolumn{1}{c|}{\begin{tabular}[c]{@{}c@{}}0.01705\\ (0.0005)\end{tabular}} & \multicolumn{1}{c|}{\begin{tabular}[c]{@{}c@{}}0.0699\\ (0.0004)\end{tabular}}  & \begin{tabular}[c]{@{}c@{}}0.0744\\ (0.0004)\end{tabular}  \\ \hline
Low pass filter                                                                          & \multicolumn{1}{c|}{\begin{tabular}[c]{@{}c@{}}0.0162\\ (0.0005)\end{tabular}}  & \multicolumn{1}{c|}{\begin{tabular}[c]{@{}c@{}}0.0696\\ (0.0018)\end{tabular}}  & \begin{tabular}[c]{@{}c@{}}0.0742 \\ (0.0016)\end{tabular} \\ \hline
                                                                                         & \multicolumn{3}{c}{Global pooling strategy}                                                                                                                                                                                    \\ \hline
Global max pooling                                                                       & \multicolumn{1}{c|}{\begin{tabular}[c]{@{}c@{}}0.0189\\ (0.0009)\end{tabular}}  & \multicolumn{1}{c|}{\begin{tabular}[c]{@{}c@{}}0.0731\\ (0.0013)\end{tabular}}  & \begin{tabular}[c]{@{}c@{}}0.0791\\ (0.0009)\end{tabular}  \\ \hline
\begin{tabular}[c]{@{}c@{}}Global max pooling\\ with global average pooling\end{tabular} & \multicolumn{1}{c|}{\begin{tabular}[c]{@{}c@{}}0.0307\\ (0.0016)\end{tabular}}  & \multicolumn{1}{c|}{\begin{tabular}[c]{@{}c@{}}0.0767\\ (0.0016)\end{tabular}}  & \begin{tabular}[c]{@{}c@{}}0.0816\\ (0.0024)\end{tabular}  \\ \hline
Generalized mean pooling                                                                 & \multicolumn{1}{c|}{\begin{tabular}[c]{@{}c@{}}0.0162\\ (0.0005)\end{tabular}}  & \multicolumn{1}{c|}{\begin{tabular}[c]{@{}c@{}}0.0696\\ (0.0018)\end{tabular}}  & \begin{tabular}[c]{@{}c@{}}0.0742 \\ (0.0016)\end{tabular} \\ \hline
\end{tabular}
\label{tbl:ab_error}
\end{table}

\begin{figure}
\subcaptionbox{\footnotesize{}}{\includegraphics[width=0.48\textwidth]{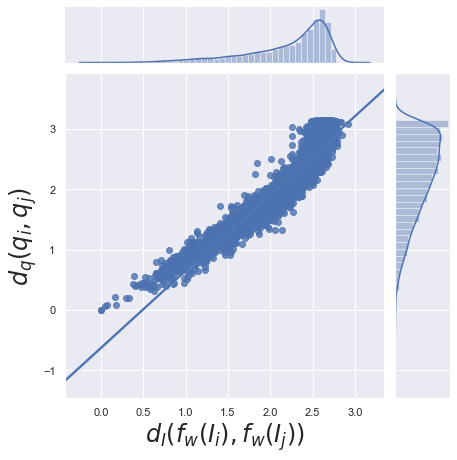}}%
\subcaptionbox{\footnotesize{}}{\includegraphics[width=0.48\textwidth]{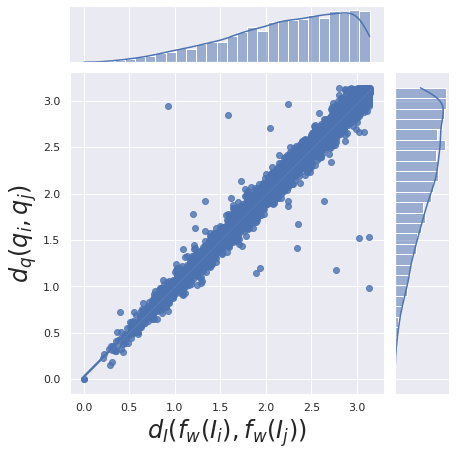}}
\caption{The regression plot between the true pairwise distance of unit quaternion and the predicted pairwise distance of unit quaternion. (a) The plot was obtained by using standard unit quaternion. (b) The plot was obtained by using our proposed architecture.}
\label{fig:distances}
\end{figure}

\subsection{Numerical study verifies the design choices within the framework}
We begin by comparing various rotation representations. Three different approaches to parameterizing $\SO(3)$ have been tested. The first is the standard unit quaternion, as utilized in \cite{banjac2021learning,lian2022end}, which provides a discontinuous representation. The second is a 6D continuous representation, achieved through Gram-Schmidt Orthogonalization, which has been employed in \cite{levy2022amortized,nashed2021cryoposenet,levy2022cryoai}. The last approach is our modified QCQP representation, which is based on \cite{peretroukhin2020smooth}. As evident from \Cref{tbl:ab_error}, the modified QCQP representation demonstrates significant performance gains over the other two representations. Notably, this approach rectifies the issue reported in \cite{banjac2021learning}, where $f_w$ was observed to underestimate the distances between image pairs with larger distances, as depicted in \Cref{fig:distances}.

Next, we turn our attention to evaluating different training styles for $f_w$. We contrast the conventional training methodology, which deploys a single CNN as seen in \cite{lian2022end,nashed2021cryoposenet,levy2022cryoai} \footnote{The CNN is trained using Adam with One cycle learning for 1,300 epochs to ensure its convergence.}, against the strategy employing a Siamese network. Additionally, our proposed training technique, which incorporates the Siamese network along with an auxiliary loss, is also assessed. \Cref{tbl:ab_error} shows that our training method significantly surpasses the other two approaches.

Regarding the blurring layer, three configurations are devised to examine the impact of the blurring layer. These configurations include an architecture devoid of a blurring layer, one with a Gaussian filter layer as described in \cite{levy2022cryoai}, and another with a low pass filter layer. \Cref{tbl:ab_error} displays the corresponding errors. We note a slight reduction in error when incorporating the blurring layer. This can be ascribed to the hypothesis that low-resolution information can facilitate the estimation of orientations in the early stages.

We also evaluate various global pooling strategies, which are used to ensure size invariance. The first strategy under consideration is global max pooling, which has been employed in \cite{lian2022end,nashed2021cryoposenet,levy2022cryoai}. The second strategy is one that is used in the fastai \cite{howard2020fastai} library, which involves concatenating the results obtained from global max pooling and global average pooling. The final strategy is generalized mean pooling, which is the approach we have adopted for this study. As can be observed from \Cref{tbl:ab_error}, the generalized mean pooling offers a slight performance enhancement compared to the other two widely-used approaches.

Regarding the effect of noise, we also measure the estimation error under different SNR levels, as shown in \Cref{tbl:snr_error}. From the table, it is clear that the SNR of particle images has a significant impact on the accuracy of estimation, which suggests that a better filtering scheme or denoising methods may improve the results.

\begin{figure}[h]
\centering
\subcaptionbox{\footnotesize{}}{\includegraphics[width=0.32\textwidth]{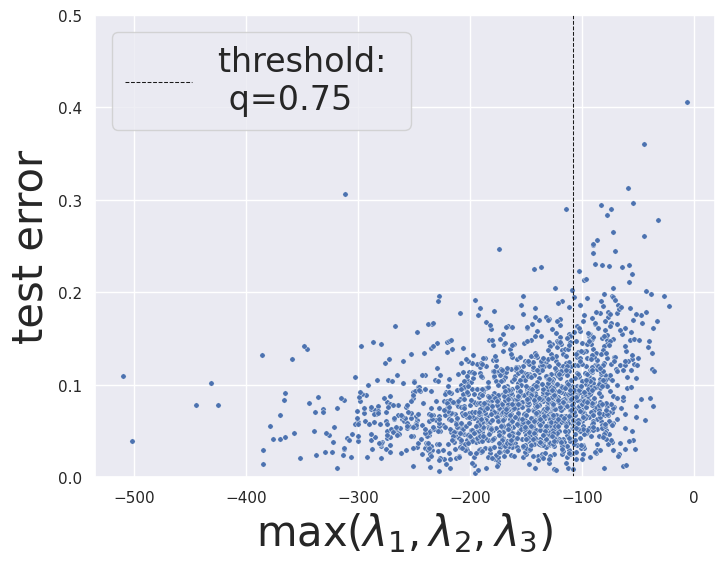}}%
\hfill 
\subcaptionbox{\footnotesize{}}{\includegraphics[width=0.32\textwidth]{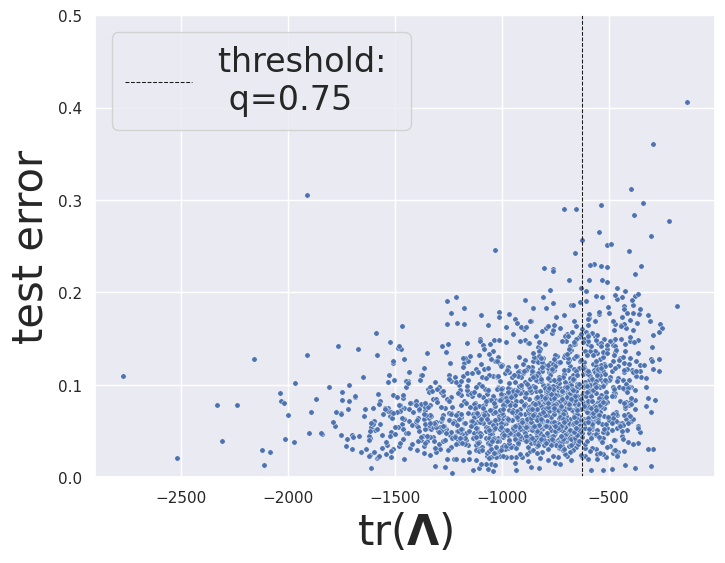}}
\hfill
\subcaptionbox{\footnotesize{}}{\includegraphics[width=0.32\textwidth]{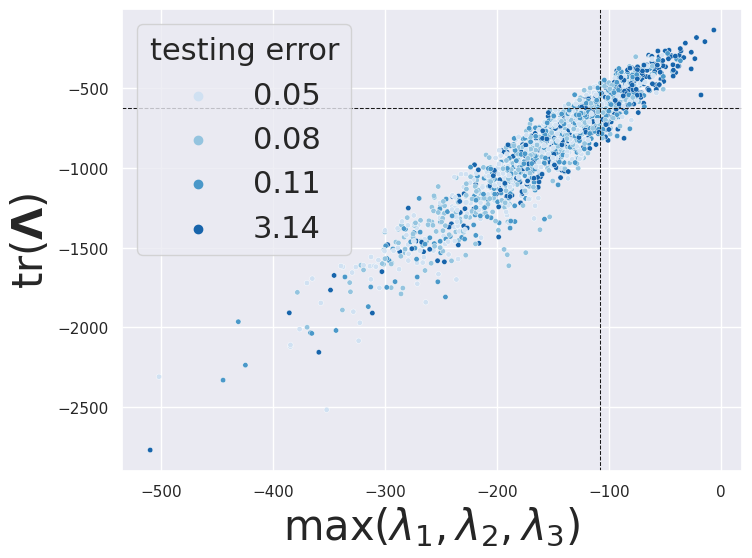}}

\subcaptionbox{\footnotesize{}}{\includegraphics[width=0.27\textwidth]{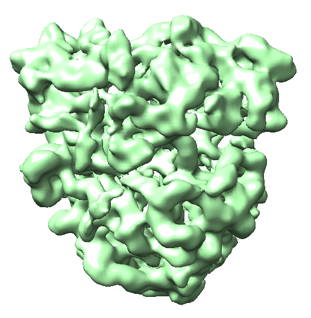}}%
\hfill 
\subcaptionbox{\footnotesize{}}{\includegraphics[width=0.25\textwidth]{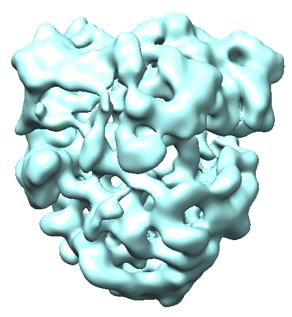}}
\hfill
\subcaptionbox{\footnotesize{}}{\includegraphics[width=0.25\textwidth]{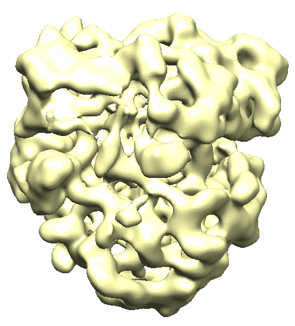}}

\caption{The scatter plot between (a) statistic \eqnref{max_lambda} and testing error, (b) statistic \eqnref{trace} and testing error and (c) statistic \eqnref{max_lambda} and \eqnref{trace} (the gradient of color is labeled by the testing error in different quantiles). 3D Reconstruction from particles with true orientations (d), estimated orientation (e) and estimated orientation after filtering (f) on the test set of synthetic 70S ribosome dataset.}
\label{fig:70_syn}
\end{figure}

\begin{figure}[h]
\centering
\subcaptionbox{\footnotesize{}}{\includegraphics[width=0.32\textwidth]{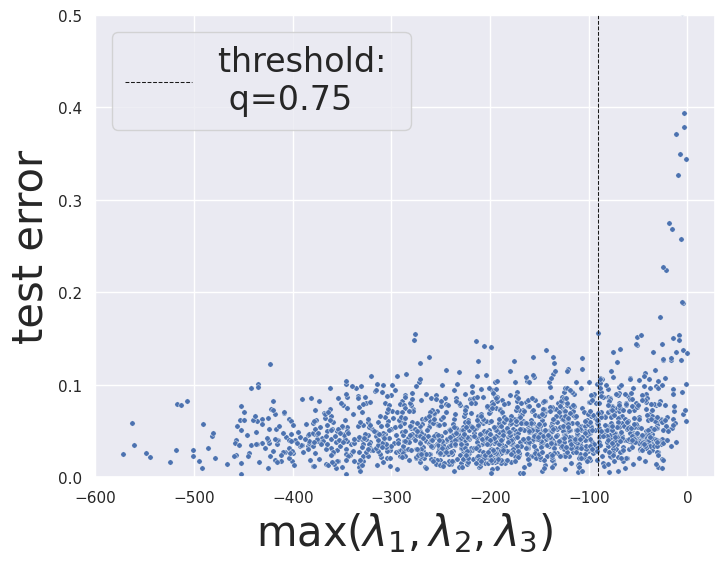}}%
\hfill
\subcaptionbox{\footnotesize{}}{\includegraphics[width=0.32\textwidth]{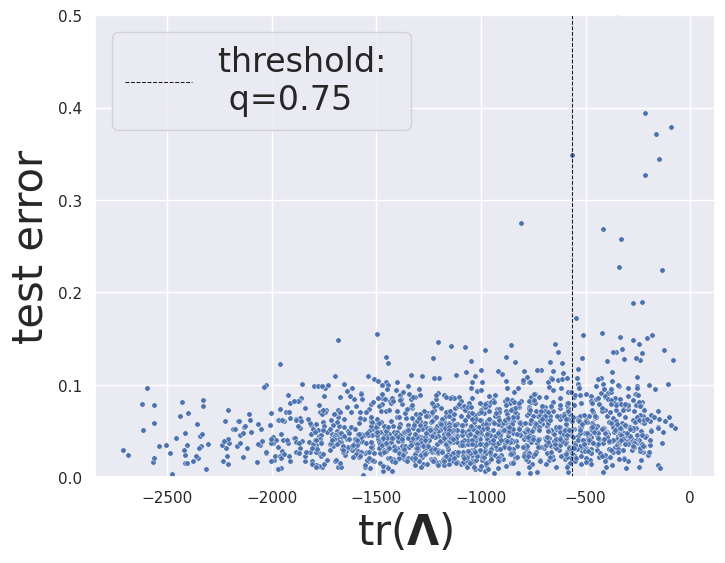}}
\hfill
\subcaptionbox{\footnotesize{}}{\includegraphics[width=0.32\textwidth]{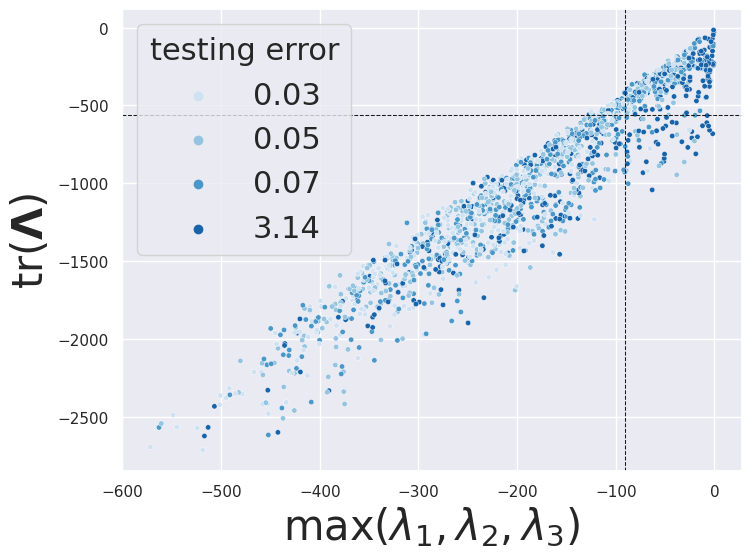}}

\subcaptionbox{\footnotesize{}}{\includegraphics[width=0.27\textwidth]{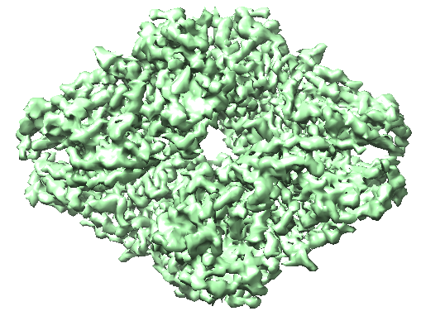}}%
\hfill 
\subcaptionbox{\footnotesize{}}{\includegraphics[width=0.25\textwidth]{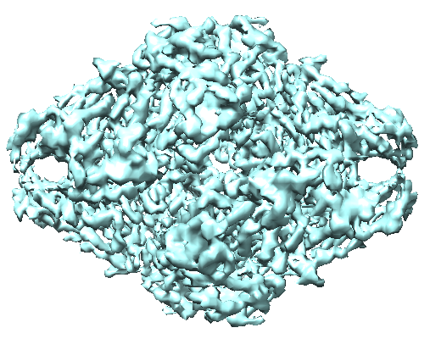}}
\hfill
\subcaptionbox{\footnotesize{}}{\includegraphics[width=0.25\textwidth]{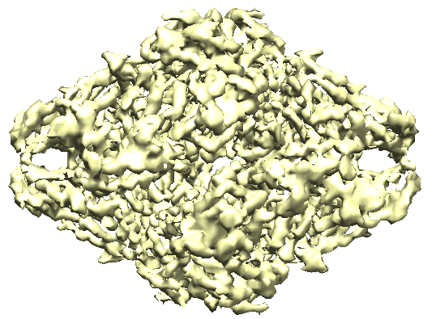}}

\caption{The scatter plot between (a) statistic \eqnref{max_lambda} and testing error, (b) statistic \eqnref{trace} and testing error and (c) statistic \eqnref{max_lambda} and \eqnref{trace} (the gradient of color is labeled by the testing error in different quantiles). 3D Reconstruction from particles with true orientations (d), estimated orientation (e) and estimated orientation after filtering (f) on the test set of synthetic beta-galactosidase dataset.}
\label{fig:beta_syn}
\end{figure}

\begin{figure}[h]
\centering
\subcaptionbox{\footnotesize{}}{\includegraphics[width=0.27\textwidth]{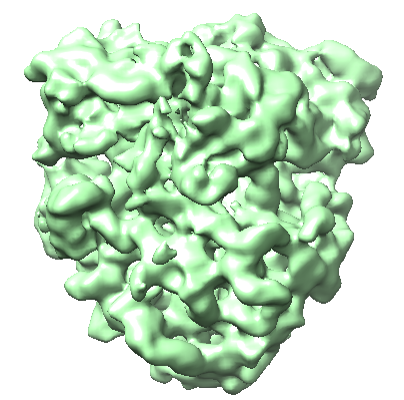}}%
\hfill 
\subcaptionbox{\footnotesize{}}{\includegraphics[width=0.25\textwidth]{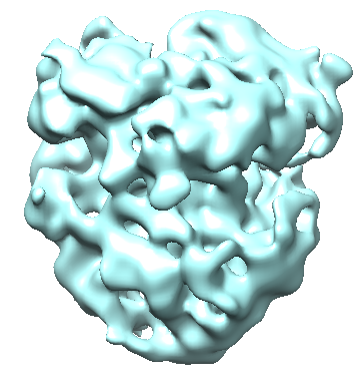}}
\hfill
\subcaptionbox{\footnotesize{}}{\includegraphics[width=0.25\textwidth]{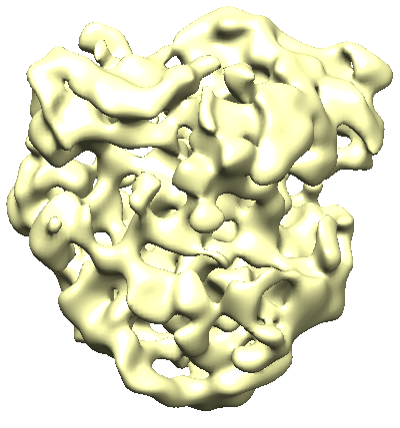}}
\caption{3D Reconstruction from particles with true orientations (a), estimated orientation (b), estimated orientation after filtering (c) on the test set of real 70S ribosome dataset.}
\label{fig:70_real}
\end{figure}

\subsection{The reconstruction pipeline of the proposed framework}
As a proof-of-concept, we attempted to address the entire inverse problem, that is, to reconstruct the density maps $\widehat{V}$  from sets of projections $\{ I_i \}$ and their respective orientations $\{ \widehat{q_i} \}$ that were obtained using the framework proposed in this study. Since we focus on the orientation estimation, we employed a direct reconstruction algorithm for the reconstruction process \footnote{Specifically, we take the shifts from the synthetic process and use the reconstruction program from cryoSPARC.}.

\Cref{fig:70_syn} illustrates the 3D structure derived from the orientation recovery accomplished by our framework, which was estimated based on the noisy training projections. With a median error of $0.07$ radians in the recovered orientations, the reconstruction utilizing the noisy testing images achieved a resolution of 14.4 \AA{}. This resolution is determined by measuring the Fourier Shell Correlation (FSC) at a self-consistency value of $0.143$, as shown in \Cref{fig:70_syn}(e). Moreover, \Cref{fig:70_syn} displays the uncertainty measures used in our framework. As can be observed from \Cref{fig:70_syn}(a)-(c), both \eqnref{max_lambda} and \eqnref{trace} serve as reasonably effective indicators of the testing error. A larger value of the statistics often corresponds to a larger testing error. As a result, we further excluded the $25\%$ of images with the highest values and used the remaining images for reconstruction, achieving the 3D reconstruction shown in \Cref{fig:70_syn}(f). This led to a slight improvement in the FSC, bringing it to 13.1 \AA{} when we use \eqnref{trace} as our filtering criteria. Notably, the corresponding value for the reconstructed volume with ground truth orientation is 11.2 \AA{}, as shown in \Cref{fig:70_syn}(d) for this dataset. These results suggest that a reasonably accurate structural reconstruction can be obtained from projections whose orientations have been determined through our framework.

\subsection{Application of the framework to the dataset with symmetry particles}
To demonstrate the application of our framework to proteins with symmetry, we downloaded the beta-galactosidase dataset, first reported in \cite{scheres2015semi}, which possesses D2 symmetry. When learning distances, two projections could be identical without originating from the same orientation if we don't account for symmetry. Our framework addresses this issue by restricting the orientations used for training. For D2 symmetry, we limit the orientation to $\bth = (\psi,\theta,\phi) \in [0,2\pi) \times [0,\pi/2) \times [0,\pi)$, ensuring a one-to-one mapping between particles and orientations. This constraint also enhances the training as the network needs to learn a smaller geometry of orientations compared to asymmetric particles. We generated the particle stack by resizing the downloaded map to a box size of 128 pixels, corresponding to a pixel size of 2.49 \AA{}. The synthetic particle stack, created using the same procedure as in the previous subsection, contains 5,000 images with an SNR of 0.1.

\Cref{fig:beta_syn} presents the results obtained by our framework when applied to this dataset. Our framework successfully recovered the orientations with a median error of $0.05$ radians. The reconstruction, conducted using noisy testing images, reached a resolution of 5.2 \AA{} as depicted in \Cref{fig:beta_syn}(e). \Cref{fig:beta_syn} also displays the uncertainty measures utilized in our framework. As shown in panels (a) to (c) of \Cref{fig:beta_syn}, \eqnref{max_lambda} and \eqnref{trace} once again serve as reasonably effective indicators of the testing error. Notably, the testing error recorded is lower than that observed in datasets containing asymmetric proteins, such as ribosomes in the previous subsection. This is expected as the higher symmetry reduces the complexity of the geometry that the network has to learn within our framework. After further excluding the $25\%$ of images with the highest values according to \eqnref{max_lambda} and using the remaining images for reconstruction, we achieved a 3D reconstruction as shown in \Cref{fig:beta_syn}(f). The FSC value remains unchanged, as the reconstructed volume with the ground truth orientation also measures 5.2 \AA{}, as depicted in \Cref{fig:beta_syn}(d) for this dataset. These results suggest that our framework can produce highly precise structural reconstructions when applied to symmetry proteins. In closing, it is noteworthy that employing the statistics of \eqnref{max_lambda} as our metric results in a marginally reduced error compared to using the statistics of \eqnref{trace}. However, the opposite trend was observed in the previous dataset. From practical experience, we conclude that both statistics are effective.

\subsection{Application of the framework to real dataset}
In order to validate the effectiveness of our approach with real data, we imported the first 5,000 images from the \href{https://www.ebi.ac.uk/pdbe/emdb/test_data.html}{E. coli 70S ribosome} dataset into our framework. To expedite the training process and focus on orientation estimation, the images were initially cropped to a resolution of $128 \times 128$ pixels with a pixel size of 2.86 \AA{}. This was followed by an alignment procedure using XMIPP \cite{strelak2021advances} for rough centering. Subsequently, we used cryoSPARC to generate a 3D density map from these particle images, which facilitated the extraction of orientations. These orientations were used as the ground truth for the training and validation splits. To mitigate potential orientation bias within the real dataset, we used stratified sampling instead of random sampling. All other parameters were retained from the previous experiment.

\Cref{fig:70_real} showcases the results obtained using experimental projections. The reconstruction achieved a resolution of 17.1 \AA{}, as determined by the Fourier Shell Correlation (FSC) shown in \Cref{fig:70_real}(b). By excluding the top $25\%$ of images based on the highest values of \eqnref{trace} and using the remaining images for reconstruction, we were able to achieve a 3D reconstruction, as depicted in \Cref{fig:70_syn}(c). This selective exclusion process improved the FSC measure, yielding a resolution of 15.9 \AA{}. These results are in line with findings from the synthetic dataset, reinforcing the fact that our framework is capable of generating reasonably accurate structural reconstructions. For reference, the reconstructed volume with the ground truth orientation has a resolution of 11.5 \AA{}, as shown in \Cref{fig:70_real}(a).

\section{Discussion and Conclusion} \label{discussion}
In this research, we delved into the application of distances between pairs of 2D cryo-EM projections of 3D protein structures as supplementary loss to enhance the learning of 3D orientations. Our framework integrates state-of-the-art components, and we methodically evaluated the contributions of each component via a numerical study. Additionally, we introduced a novel architecture incorporating a blurring layer to increase accuracy and a modified QCQP layer for uncertainty quantification. Evaluations conducted on synthetic datasets provided critical insights into the performance of the proposed strategy. 

Key findings in this study include the framework's proficiency in generating more accurate estimations compared to prior architectures, addressing biased distance learning as discussed in \cite{banjac2021learning} without employing the 3D volume in the learning phase. Moreover, our methodology proficiently recovered orientations with an error margin of $0.07$ and $0.05$ radians in the test set for asymmetric and symmetric proteins, respectively. Utilizing downstream reconstruction algorithms, our framework can deliver precise initial and final models for asymmetric and symmetric proteins, respectively. Additionally, our framework is adaptable to real datasets with outcomes in agreement with synthetic data. Importantly, the uncertainty metric proves valuable as an indicator of testing error, with high-resolution results achievable by excluding particle images with greater uncertainty.

Our framework bears the potential to augment extant research. For instance, studies as in \cite{banjac2021learning,lian2022end} could incorporate our innovative layers, sampling schemes, or rotational representations to improve rotation estimations. Likewise, generative models employing autoencoder architectures like \cite{levy2022amortized,nashed2021cryoposenet,levy2022cryoai} could bolster their encoder components through our framework, resulting in enhanced reconstructions or more insightful conformational analysis. Notably, prior autoencoder frameworks addressing end-to-end homogeneous and heterogeneous reconstruction did not prioritize encoder design. Our findings can fill the gap in the design consideration of the encoder. Our framework is structured into a modular package, {\it cryo-forum}, designed for accessibility to the cryo-EM community. The framework offers developers the versatility to choose among different rotation representations, global pooling methods, and training modalities. Users can also opt to include the preprocessing layer and have the convenience of customizing the architecture, such as omitting the final two convolutional layers for expedited training.

Furthermore, the incorporation of uncertainty metrics offers avenues for creating an efficient reconstruction pipeline for real datasets as follows: Initially, particle pickers like Topaz \cite{bepler2019positive} or crYOLO \cite{wagner2019sphire} can be deployed for automated particle selection from micrographs using a higher confidence threshold. These particles can be input into 3D refinement frameworks such as cryoSPARC or RELION for orientation determination, serving as training data for developing a reliable orientation estimator in our framework. A second round of particle picking with a lower confidence threshold can then curate a larger dataset, with orientations inference via the trained model. Our uncertainty metrics can be used to filter out images with high uncertainty, retaining the rest for reconstruction. Moreover, our framework's orientation outputs can initialize parameters in software like cryoSPARC or RELION for further optimization. This approach fosters an effective and robust reconstruction pipeline that combines automated particle picking, orientation estimation, and uncertainty-driven filtering with conventional refinement techniques. Another potential usage of the uncertainty metric is incorporating it into calculating the reconstruction process for weighted back projection to obtain more accurate 3D results.

Finally, future work could benefit from training the framework on a diverse range of cryo-EM datasets to develop a pre-trained model for transfer learning. In this regard, generating realistic cryo-EM projections may be achieved by adopting a more realistic representation of cryo-EM physics \cite{himes2021cryo,vulovic2013image,rullgaard2011simulation} or utilizing the EMPIAR \cite{iudin2016empiar}. The final step entails rigorous validation of novel proteins, specifically proteins not previously exposed to the framework in simulated projections. The SNN's inherent ability to predict relationships between projections, enabling the abstraction of specific volumes, is a promising feature. The structural similarities among proteins should bolster learning, paving the way for creating a pre-trained model.

\section*{Data availability}
The two synthetic datasets were generated using RELION and are available at the Github repository \href{https://github.com/phonchi/Cryo-forum/tree/main}{https://github.com/phonchi/Cryo-forum/tree/main}. The 70S ribosome experimental dataset was downloaded from \href{https://www.ebi.ac.uk/emdb/test_data.html}{https://www.ebi.ac.uk/emdb/test\_data.html}.

\section*{Code availability}
The software of our framework and analysis scripts are implemented in custom software and will be available at \href{https://github.com/phonchi/Cryo-forum/tree/main}{https://github.com/phonchi/Cryo-forum/tree/main}.

\section*{Acknowledgments}
This work was supported by [MOST 110-2118-M-110-003-MY2] to S.-Z.C.. 

\section*{Author contributions}
S.-Z.C. developed the code, performed the experiments and wrote the manuscript.

\subsection*{Competing interests}
The authors declare no competing interests.


%
\bibliography{ref}

\newpage
\appendix

\renewcommand{\figurename}{Supplementary Figure}
\renewcommand{\tablename}{Supplementary Table}
\crefalias{figure}{appendixfigure}
\crefalias{table}{appendixtable}

\setcounter{figure}{0}  
\setcounter{table}{0} 
\setcounter{page}{1}


\section{Supplementary Figures}

\begin{figure}[ht!]
\subcaptionbox{\footnotesize{}}{\includegraphics[width=0.48\textwidth]{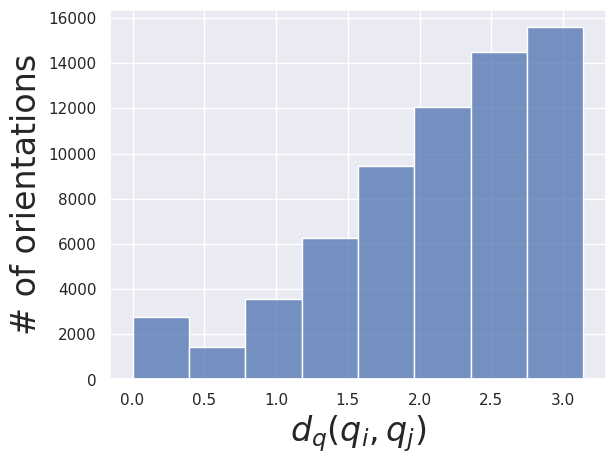}}%
\subcaptionbox{\footnotesize{}}{\includegraphics[width=0.48\textwidth]{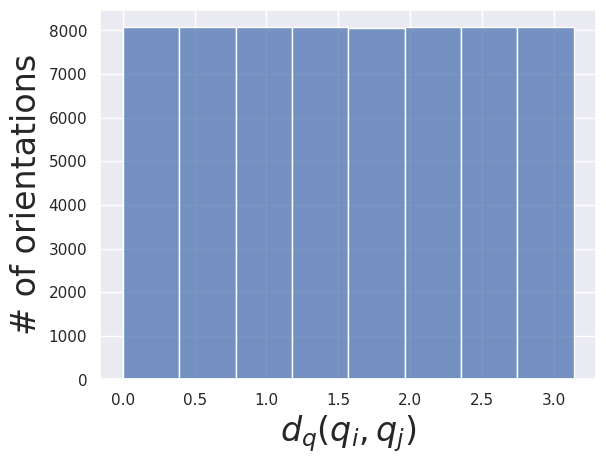}}
\caption{The histogram of the distances between quaternions with (a) Random sampling. (b) Stratified sampling. The number of bins is set to 8.}
\label{fig:histograms}
\end{figure}

\begin{figure}[h]
\centering
    \includegraphics[width=0.9\textwidth]{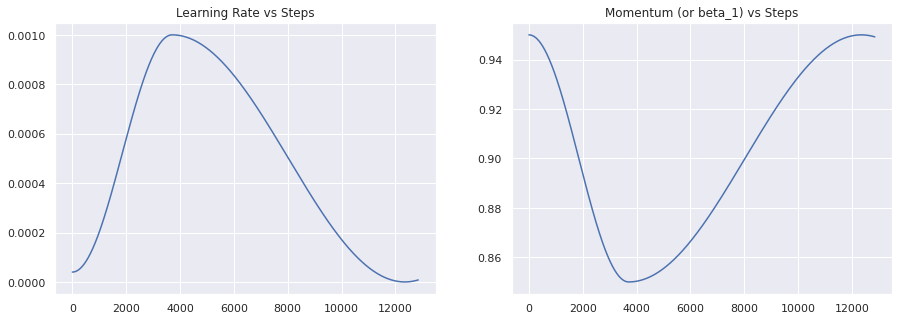}
\caption{The learning rate and momentum scheduling in our framework.}
\label{fig:1cycle}
\end{figure}


\begin{table}[]
\small
\captionsetup{justification=centering, singlelinecheck=false}
\centering
\caption{The network structure of our proposed framework. Dropout rate is set to 0.3 and $l_2$ regularization with a factor of 0.001 is used for the weights.}
\begin{tabular}{@{}cccccc@{}}
\toprule
Layer name                                                                               & \begin{tabular}[c]{@{}c@{}} Kernel size\\ and structure\end{tabular}                               & Output Size                           & Layer name                                                                              & \begin{tabular}[c]{@{}c@{}} Kernel size\\ and structure\end{tabular}                               & Output Size    \\ \midrule
\multicolumn{1}{c|}{\begin{tabular}[c]{@{}c@{}}Input image \\ (Gray scale)\end{tabular}} & \multicolumn{1}{c|}{}                                                                                   & \multicolumn{1}{c|}{$(128, 128, 1)$}  & \multicolumn{1}{c|}{Conv7}                                                              & \multicolumn{1}{c|}{\begin{tabular}[c]{@{}c@{}}$3 \times 3$, stride 1,\\ BN, PReLU\end{tabular}} & $(8, 8, 512)$  \\ \midrule
\multicolumn{1}{c|}{Blurring}                                                      & \multicolumn{1}{c|}{}                                                                                   & \multicolumn{1}{c|}{$(128, 128, 6)$} & \multicolumn{1}{c|}{Conv8}                                                              & \multicolumn{1}{c|}{\begin{tabular}[c]{@{}c@{}}$3 \times 3$, stride 1,\\ BN, PReLU\end{tabular}} & $(8, 8, 512)$  \\ \midrule
\multicolumn{1}{c|}{Conv1}                                                               & \multicolumn{1}{c|}{\begin{tabular}[c]{@{}c@{}}$7 \times 7$, stride 1,\\ BN, PReLU\end{tabular}} & \multicolumn{1}{c|}{$(128, 128, 32)$} & \multicolumn{1}{c|}{MaxPool5}                                                           & \multicolumn{1}{c|}{\begin{tabular}[c]{@{}c@{}}$2 \times 2$, stride 2,\\ Dropout layer\end{tabular}}    & $(4, 4, 512)$  \\ \midrule
\multicolumn{1}{c|}{MaxPool1}                                                            & \multicolumn{1}{c|}{\begin{tabular}[c]{@{}c@{}}$2 \times 2$, stride 2,\\ Dropout layer\end{tabular}}    & \multicolumn{1}{c|}{$(64, 64, 32)$}   & \multicolumn{1}{c|}{Conv9}                                                              & \multicolumn{1}{c|}{\begin{tabular}[c]{@{}c@{}}$3 \times 3$, stride 1,\\ BN, PReLU\end{tabular}} & $(4, 4, 1024)$ \\ \midrule
\multicolumn{1}{c|}{Conv2}                                                               & \multicolumn{1}{c|}{\begin{tabular}[c]{@{}c@{}}$5 \times 5$, stride 1,\\ BN, PReLU\end{tabular}} & \multicolumn{1}{c|}{$(64, 64, 64)$}   & \multicolumn{1}{c|}{Conv10}                                                             & \multicolumn{1}{c|}{\begin{tabular}[c]{@{}c@{}}$3 \times 3$, stride 1,\\ BN, PReLU\end{tabular}} & $(4, 4, 1024)$ \\ \midrule
\multicolumn{1}{c|}{MaxPool2}                                                            & \multicolumn{1}{c|}{\begin{tabular}[c]{@{}c@{}}$2 \times 2$, stride 2,\\ Dropout layer\end{tabular}}    & \multicolumn{1}{c|}{$(32, 32, 64)$}   & \multicolumn{1}{c|}{MaxPool6}                                                           & \multicolumn{1}{c|}{\begin{tabular}[c]{@{}c@{}}$2 \times 2$, stride 2,\\ Dropout layer\end{tabular}}    & $(2, 2, 1024)$ \\ \midrule
\multicolumn{1}{c|}{Conv3}                                                               & \multicolumn{1}{c|}{\begin{tabular}[c]{@{}c@{}}$3 \times 3$, stride 1,\\ BN, PReLU\end{tabular}} & \multicolumn{1}{c|}{$(32, 32, 128)$}  & \multicolumn{1}{c|}{\begin{tabular}[c]{@{}c@{}}Generalize \\ mean pooling\end{tabular}} & \multicolumn{1}{c|}{}                                                                                   & $1024$         \\ \midrule
\multicolumn{1}{c|}{Conv4}                                                               & \multicolumn{1}{c|}{\begin{tabular}[c]{@{}c@{}}$3 \times 3$, stride 1,\\ BN, PReLU\end{tabular}} & \multicolumn{1}{c|}{$(32, 32, 128)$}  & \multicolumn{1}{c|}{Dense}                                                              & \multicolumn{1}{c|}{}                                                                                   & $10$           \\ \midrule
\multicolumn{1}{c|}{MaxPool3}                                                            & \multicolumn{1}{c|}{\begin{tabular}[c]{@{}c@{}}$2 \times 2$, stride 2,\\ Dropout layer\end{tabular}}    & \multicolumn{1}{c|}{$(16, 16, 128)$}  & \multicolumn{1}{c|}{\begin{tabular}[c]{@{}c@{}}Modified \\ QCQP\end{tabular}}           & \multicolumn{1}{c|}{}                                                                                   & $4$            \\ \midrule
\multicolumn{1}{c|}{Conv5}                                                               & \multicolumn{1}{c|}{\begin{tabular}[c]{@{}c@{}}$3 \times 3$, stride 1,\\ BN, PReLU\end{tabular}} & \multicolumn{1}{c|}{$(16, 16, 256)$}  &                                                                                         &                                                                                                         &                \\ \cmidrule(r){1-3}
\multicolumn{1}{c|}{Conv6}                                                               & \multicolumn{1}{c|}{\begin{tabular}[c]{@{}c@{}}$3 \times 3$, stride 1,\\ BN, PReLU\end{tabular}} & \multicolumn{1}{c|}{$(16, 16, 256)$}  &                                                                                         &                                                                                                         &                \\ \cmidrule(r){1-3}
\multicolumn{1}{c|}{MaxPool4}                                                            & \multicolumn{1}{c|}{\begin{tabular}[c]{@{}c@{}}$2 \times 2$, stride 2,\\ Dropout layer\end{tabular}}    & \multicolumn{1}{c|}{$(8, 8, 256)$}    &                                                                                         &                                                                                                         &                \\ \cmidrule(r){1-3}
\end{tabular}
\label{tbl:arch}
\end{table}

\begin{figure}[h]
\centering
    \includegraphics[width=0.9\textwidth]{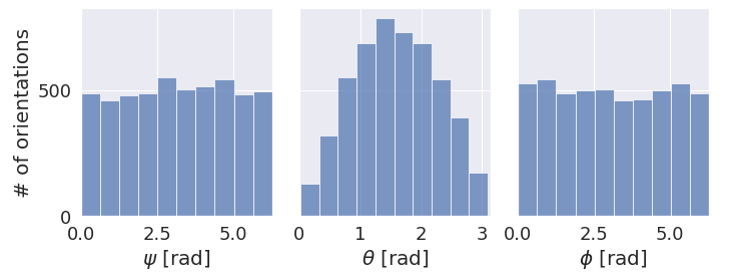}
\caption{Orientation sampling according to the HEALPix framework of RELION \cite{scheres2012relion}.}
\label{fig:sampling}
\end{figure}

\begin{table}[]
\small
\captionsetup{justification=centering, singlelinecheck=false}
\centering
\caption{The testing error associate with different SNR levels.}
\begin{tabular}{@{}c|c@{}}
\toprule
SNR   & Testing error \\ \midrule
Clean & 0.0221        \\ \midrule
1.0   & 0.0324        \\ \midrule
0.7   & 0.0349        \\ \midrule
0.4   & 0.0393        \\ \midrule
0.2   & 0.0507        \\ \midrule
0.1   & 0.0742        \\ \bottomrule
\end{tabular}
\label{tbl:snr_error}
\end{table}

\end{document}